\documentclass[10pt,a4paper,twoside]{article}

\usepackage{vmargin,fancyheadings,multicol,multirow,ifthen,cite,
            graphicx,wrapfig,calc,dcolumn,apalike,setspace,
            boxedminipage,rotating,textcomp,picinpar,longtable,
            url,amsmath,amssymb,ogonek} 

\usepackage[tight]{subfigure}
\usepackage{imc2018}

\begin{document}
\SetPaperBodyFont

\begin{IMCpaper}{
\title{Balloon-borne video observations of Geminids 2016}
\author{Francisco~Oca\~{n}a$^1$, Alejandro~S\'{anchez}~de~Miguel$^{1,2,3}$, ORISON team, and Daedalus Project.
        \thanks{$^1\,$Dpto. de F\`{ı}sica de la Tierra y Astrof\`{ı}sica, Facultad de Ciencias F\`{ı}sicas, Universidad Complutense de Madrid, E-28040 Madrid,
Spain\\
                     \texttt{focana@sciops.esa.int}\\[1ex]
                $^2\,$Environment and Sustainability Institute, University of Exeter, Penryn, Cornwall TR10 9FE, UK,\\ 
                     \texttt{a.sanchez-de-miguel@exeter.ac.uk}\\[1ex]
                $^3\,$Instituto de Astrofísica de Andalucía, Glorieta de la Astronomía, s/n, C.P.18008 Granada, Spain}}%
\abstract{We investigate the observation of meteors with video cameras in stratospheric balloons, overcoming tropospheric handicaps like weather and extinction. We have studied the practical implementation of the idea, designed and tested instrumentation for balloon-borne missions. We have analysed the data of the Geminids 2016 campaign, determining the meteoroid flux just before the maximum.}}%
\index{Oca\~{n}a F.}%
\index{S\'{anchez}~de~Miguel A.}%
\vspace*{-3\baselineskip}

\section{Introduction}
This text is an adaption of the work by the first author for his PhD Thesis: Techniques for near-Earth interplanetary matter detection and characterisation from optical ground-based observatories \cite{2017PhDT........58O}. Refer to his thesis for further detail. The lines here are a summary of the presentation given, for the sake of completeness of these proceedings of the IMC 2018 in Pezinok-Modra.
The multimedia material shown during the presentation at IMC 2018 can be found in the Zenodo repository for the ORISON Project and Daedalus Project. Zenodo is an open-access repository aimed for datasets, it provides them a Data Object Identifier (DOI) and it is intended for otherwise orphan records, making them easier to cite. All our datasets are licensed under Creative Commons Attribution 4.0 [CC BY 4.0] (i.e., you are free to share, adapt, use or whatever, just give appropriate credit). Moreover the authors encourage you to use the data and contact them for more details. 

These are the DOI of the three datasets:

\begin{itemize}
	\item Geminids 2016 - Part 1 :\\ https://doi.org/10.5281/zenodo.579708 \\ \cite{sanchez_de_miguel_alejandro_2017_579708}
	\item Geminids 2016 - Part 2 :\\ https://doi.org/10.5281/zenodo.801598 \\ \cite{sanchez_de_miguel_alejandro_2017_801598}
	\item Geminids 2016 - Part 3 :\\ https://doi.org/10.5281/zenodo.842269 \\ \cite{sanchez_de_miguel_alejandro_2017_842269}
\end{itemize}

\section{Balloon-borne meteoroid flux determination: Geminids 2016 show-case}

In order to determine meteoroid fluxes, the area surveyed by a sensor is calculated as the projection of the field of view onto the meteor layer. The method employed here is a generalization of the method defined by Koschack and Rendtel (1990) and later applied for cameras by Bellot Rubio (1994b). The work by the first author \cite{2017PhDT........58O} includes the height of the observer, $ h_b$, as an extension for airborne and balloon-borne observations. Measurements are limited to horizon elevation due to the difficulties to correct for extinction.
 
To perform the observations of Geminids, we used the balloonborne platform developed by Daedalus project. Since 2011, when we sent a low-light video camera to observe the Draconids 2011 outburst \cite{ocana2013first}, they carried a meteor detection payload for 8 nighttime missions so far. Since 2016 the payloads have flown on-board the ORISON Pathfinder missions. ORISON is a H2020 project to study feasibility of innovative astronomical research infrastructure based on stratospheric balloons (López-Moreno et al., 2016).

ORISON project provided new hardware, and the B\&W low-light 1/2'' CCD video camera was replaced by a full-frame colour CMOS videocamera with better sensitivity, a Sony~$\alpha$7S (formally Sony ILCE-A7S). The colour information provides basic spectral information from the objects \cite{ocanapimo}. The Sony $\alpha$7S is an EVIL (Electronic Viewfinder with Interchangeable Lens) camera with a backlit CMOS sensor and a self-recording system included. The camera hosts a full-frame format (35.6~mm~$\times$~23.8~mm Exmor\textsuperscript{TM}) CMOS sensor with a total of 12.2 megapixels and a pixel size of 8.4~$\mu$m and a sensitivity up to 409600 ISO. The nominal configuration we use for the record of meteors is full-HD-1080p colour frames (1920 x 1980 pixels = 2 Mpix), at 30 fps with $\frac{1}{50}$~s exposure time and sensitivity of 60000 ISO (over that value the noise increases with no apparent increase of sensitivity). The video is stored in clips of 1m44s with a data rate of 50 Mbps, and a total filesize of 650~MB. The format of the video is MPEG-4 Part 14 (international standard ISO/IEC 14496-14:2003), using the Sony proprietary codec for professional videos XAVC S. 

The camera has been flown in several missions, some of them during meteor shower peaks. We have selected one as a show-case. For Geminids 2016 the lens we used is a Samyang with a focal length of 24~mm and f/1.5, that produces a slight vignetting in the corners of the full format chip.

We have analysed the astrometry of the image splitting the video in all the frames, and each frame in the R, G, B channels. Using the suite \texttt{astrometry.net} \cite{lang2010astrometry} we calculated the plate constants. The plate scale of the system in video mode is 153 arcsec/pixel. Measured PSF for the stars in the FoV has a FWHM of 460~arcsec on average, in the range of a critical sampling frequency according to the Nyquist theorem. The field of view is $82\,^{\circ} \times 46\,^{\circ}$ with the centre at $hf=0\,^{\circ}$.

The launch took place at 23h17m UT the 13\textsuperscript{th} December 2016 and the burst took place at 01h50m~UT. In total the mission lasted 276 minutes as the probe landed at 03h55m~UT. The first visual inspection of the video shows stable and unstable phases. We have used the date from the 3-axis accelerators to identify these phases using the Lomb method for frequency analysis \cite{lomb1976least,ruf1999lomb}. During the whole mission we find a constant 1-second vertical tilt and 3 to 8-second roll movement (partial rotation along the optical axis). The other movement present is the rotation around the vertical axis of the probe, that results in a panning movement of the camera along the whole range of azimuth. The period of this movement changes during the different phases, and we define the stable phases when the period is larger than 6 minutes, that corresponds to a movement of $1^{\circ}$/s. This selection yields a useful period of 5 consecutive clips, for a total of 8 minutes and 40 seconds, from 01h40m UT on.

The video is analysed by visual inspection and only the upper half of the frame is considered (elevation$>0\,^{\circ}$), yielding an effective field of view of $82\,^{\circ} \times 23\,^{\circ}$. For the meteor count we have followed the visual analysis method as described in \cite{jenniskens1999activity}. As Jenniskens states the researcher only detects 70\% $\pm$ 30\% during the first visualisation, and up to 3 viewings were needed for our video clips. Star limiting magnitude was calculated measuring frames in G band with V magnitudes from catalogue, and meteor magnitudes were derived by visual comparison with stars. The star limiting magnitude is 6.0 for the period analysed, and we get the same value for the meteor video limiting magnitude $vlm$ as the magnitude loss due to the meteor speed is $<0.1$ magnitudes thanks to slow speed of meteors when pointing to the horizon.

The area monitored was 525729~km$^2$ and the correction factors yielded an effective area $A_{red}$=26110~{km$^2$}. For the full period we get an average value of $(15~\pm~3)~\cdot~10^{-3}$ $\textrm{meteoroids}~ \textrm{km}^{-2} \textrm{h}^{-1}$ producing meteors brighter than magnitude 6.5 (meteoroid mass $>$ 0.6~mg). There are no published results available for Geminids 2016, but the shower is stable through years and we can compare with other values from literature for solar longitude $261.1^{\circ}$, which range between 15 to 60 $\textrm{meteoroids}~ \textrm{km}^{-2} \textrm{h}^{-1}$ \cite{imonet,blaauw2014characteristics,nesluvsan2015summary,blaauw2016mass}. Therefore our results for Geminids 2016 from balloon-borne observations are in agreement with the values from previous years.

\nocite{*}
\bibliographystyle{imo2}
\bibliography{2018-A00-ocana}

\end{IMCpaper}
\end{document}